\documentclass[longauth]{aa}

\usepackage{graphicx}
\usepackage{txfonts}
\usepackage{gensymb} 
\usepackage{natbib}
\bibpunct{(}{)}{;}{a}{}{,} 

\usepackage{hyperref}    
\hypersetup{colorlinks=true, linkcolor=blue, citecolor=blue, urlcolor=blue}

\newcommand{\sophi}{SO/PHI} 

\newcommand{\aia}{SDO/AIA}

\newcommand{\hrt}{SO/PHI-HRT}
\newcommand{\blos}{$B_{\mathrm{LOS}}$}
\newcommand{\eui}{EUI}

\newcommand{\hri}{EUI/HRI}
\newcommand{\euv}{$\mathrm{HRI_{EUV}}$}
\newcommand{\lya}{$\mathrm{HRI_{Lya}}$}

\newcommand{\bm}[1]{\mbox{\boldmath$#1$\unboldmath}}

\begin{document}

\title{Coronal voids and their magnetic nature}

   \author{J.D.~N\" olke\inst{1}\thanks{\hbox{Corresponding author: J.D.~N\" olke} \hbox{\email{noelke@mps.mpg.de}}}\orcid{0000-0003-1988-4494} \and
   S.K.~Solanki\inst{1}\orcid{0000-0002-3418-8449} \and
   J.~Hirzberger\inst{1} \and
   H.~Peter\inst{1}\orcid{0000-0001-9921-0937} \and
   L.P.~Chitta\inst{1}\orcid{0000-0002-9270-6785} \and
   F.~Kahil\inst{1}\orcid{0000-0002-4796-9527} \and
   G.~Valori\inst{1}\orcid{0000-0001-7809-0067} \and
   T.~Wiegelmann\inst{1}\orcid{0000-0001-6238-0721} \and
   D.~Orozco~Su\' arez\inst{2}\orcid{0000-0001-8829-1938} \and
   K.~Albert\inst{1} \and
   N. Albelo~Jorge\inst{1} \and
   T.~Appourchaux\inst{3}\orcid{0000-0002-1790-1951} \and 
   A.~Alvarez-Herrero\inst{4}\orcid{0000-0001-9228-3412} \and
   J.~Blanco Rodr\'\i guez\inst{5}\orcid{0000-0002-2055-441X} \and
   A.~Gandorfer\inst{1}\orcid{0000-0002-9972-9840} \and
   D.~Germerott\inst{1} \and
   L.~Guerrero\inst{1} \and
   P.~Gutierrez-Marques\inst{1}\orcid{0000-0003-2797-0392} \and
   M.~Kolleck\inst{1} \and
   J.C.~del~Toro~Iniesta\inst{2}\orcid{0000-0002-3387-026X} \and
   R.~Volkmer\inst{6} \and
   J.~Woch\inst{1}\orcid{0000-0001-5833-3738} \and 
   B.~Fiethe\inst{7}\orcid{0000-0002-7915-6723} \and
   J.M.~Gómez~Cama\inst{8}\orcid{0000-0003-0173-5888} \and
   I.~P\' erez-Grande\inst{9}\orcid{0000-0002-7145-2835} \and 
   E.~Sanchis~Kilders\inst{5}\orcid{0000-0002-4208-3575} \and
   M.~Balaguer~Jiménez\inst{2}~\and
   L.R.~Bellot~Rubio\inst{2}\orcid{0000-0001-8669-8857} \and
   D.~Calchetti\inst{1}\orcid{0000-0003-2755-5295} \and
   M.~Carmona\inst{8} \and
   W.~Deutsch\inst{1} \and
   A.~Feller\inst{1} \and
   G.~Fernandez-Rico\inst{1,9}\orcid{0000-0002-4792-1144} \and
   A.~Fern\' andez-Medina\inst{4} \and
   P.~Garc\'\i a~Parejo\inst{4}\orcid{0000-0003-1556-9411} \and 
   J.L.~Gasent~Blesa\inst{5}\orcid{0000-0002-1225-4177} \and 
   L.~Gizon\inst{1,10}\orcid{0000-0001-7696-8665} \and 
   B.~Grauf\inst{1} \and 
   K.~Heerlein\inst{1} \and
   A.~Korpi-Lagg\inst{1}\orcid{0000-0003-1459-7074} \and
   T.~Lange\inst{7} \and 
   A.~L\' opez Jim\' enez\inst{2} \and 
   T.~Maue\inst{6,15} \and 
   R.~Meller\inst{1} \and
   A.~Moreno Vacas\inst{2}\orcid{0000-0002-7336-0926} \and
   R.~M\" uller\inst{1} \and
   E.~Nakai\inst{6} \and 
   W.~Schmidt\inst{6} \and
   J.~Schou\inst{1}\orcid{0000-0002-2391-6156} \and
   U.~Sch\" uhle \inst{1}\orcid{0000-0001-6060-9078} \and
   J.~Sinjan \inst{1}\orcid{0000-0002-5387-636X} \and
   J.~Staub\inst{1}\orcid{0000-0001-9358-5834} \and
   H.~Strecker \inst{2}\orcid{0000-0003-1483-4535} \and 
   I.~Torralbo\inst{9}\orcid{0000-0001-9272-6439} \and 
   D.~Berghmans\inst{11} \and
   E.~Kraaikamp\inst{11} \and
   L.~Rodriguez\inst{11} \and
   C.~Verbeeck\inst{11} \and
   A.N.~Zhukov\inst{11,16}\orcid{0000-0002-2542-9810} \and
   F.~Auchere\inst{12} \and
   E.~Buchlin\inst{12} \and
   S.~Parenti\inst{12} \and
   M.~Janvier\inst{12} \and
   K.~Barczynski\inst{17,13} \and
   L.~Harra\inst{13,17} \and
   C.~Schwanitz\inst{17,13} \and
   R.~Aznar Cuadrado\inst{1} \and
   S.~Mandal\inst{1} \and
   L.~Teriaca\inst{1} \and
   D.~Long\inst{14,18}\orcid{0000-0003-3137-0277} \and
   P.~Smith\inst{14}
   }

   \institute{
         Max-Planck-Institut f\"ur Sonnensystemforschung, Justus-von-Liebig-Weg 3,
         37077 G\"ottingen, Germany \\ \email{solanki@mps.mpg.de}
         \and
         Instituto de Astrofísica de Andalucía (IAA-CSIC), Apartado de Correos 3004,
         E-18080 Granada, Spain \\ \email{jti@iaa.es}
         \and
         Univ. Paris-Sud, Institut d’Astrophysique Spatiale, UMR 8617,
         CNRS, B\^ atiment 121, 91405 Orsay Cedex, France
         \and
         Instituto Nacional de T\' ecnica Aeroespacial, Carretera de
         Ajalvir, km 4, E-28850 Torrej\' on de Ardoz, Spain
         \and
         Universitat de Val\`encia, Catedr\'atico Jos\'e Beltr\'an 2, E-46980 Paterna-Valencia, Spain
         \and
         Leibniz-Institut für Sonnenphysik, Sch\" oneckstr. 6, D-79104 Freiburg, Germany
         \and
         Institut f\"ur Datentechnik und Kommunikationsnetze der TU
         Braunschweig, Hans-Sommer-Str. 66, 38106 Braunschweig,
         Germany
         \and
         University of Barcelona, Department of Electronics, Carrer de Mart\'\i\ i Franqu\`es, 1 - 11, 08028 Barcelona, Spain
         \and
         Instituto Universitario "Ignacio da Riva", Universidad Polit\'ecnica de Madrid, IDR/UPM, Plaza Cardenal Cisneros 3, E-28040 Madrid, Spain
         \and
         Institut f\"ur Astrophysik, Georg-August-Universit\"at G\"ottingen, Friedrich-Hund-Platz 1, 37077 G\"ottingen, Germany
         \and
         %
         Solar-Terrestrial Centre of Excellence -- SIDC, Royal Observatory of Belgium, Ringlaan 3, 1180 Brussels, Belgium
         \and
         Université Paris-Saclay, CNRS, Institut d'Astrophysique Spatiale, 91405, Orsay, France
         \and
         Physikalisch-Meteorologisches Observatorium Davos, World Radiation Center, 7260, Davos Dorf, Switzerland
         \and
         University College London, Mullard Space Science Laboratory, Holmbury St. Mary, Dorking, Surrey, RH5 6NT, UK
         \and 
         Fraunhofer Institute for High-Speed Dynamics, Ernst-Mach-Institut, EMI Ernst-Zermelo-Str. 4, 79104 Freiburg, Germany
         \and
         Skobeltsyn Institute of Nuclear Physics, Moscow State University, 119992 Moscow, Russia
         \and 
         Institute for Particle Physics and Astrophysics, ETH Z\"urich, 8093 Zurich, Switzerland
         \and
         Astrophysics Research Centre, School of Mathematics and Physics, Queen’s University Belfast, University Road, Belfast, BT7 1NN, Northern Ireland, UK
         }
   
   \date{[Received 2023 January 31, accepted 2023 August 31]}

\abstract
   {Extreme ultraviolet (EUV) observations of the quiet solar atmosphere reveal extended regions of weak emission compared to the ambient quiescent corona. The magnetic nature of these coronal features is not well understood.}
   {We study the magnetic properties of the weakly emitting extended regions, which we name coronal voids. In particular, we aim to understand whether these voids result from a reduced heat input into the corona or if they are associated with mainly unipolar and possibly open magnetic fields, similar to coronal holes.}
   {We defined the coronal voids via an intensity threshold of 75\% of the mean quiet-Sun (QS) EUV intensity observed by the high-resolution EUV channel (\euv) of the Extreme Ultraviolet Imager on Solar Orbiter. The line-of-sight magnetograms of the same solar region recorded by the High Resolution Telescope of the Polarimetric and Helioseismic Imager allowed us to compare the photospheric magnetic field beneath the coronal voids with that in other parts of the QS.}
   {The coronal voids studied here range in size from a few granules to a few supergranules and on average exhibit a reduced intensity of 67\% of the mean value of the entire field of view. The magnetic flux density in the photosphere below the voids is 76\% (or more) lower than in the surrounding QS. Specifically, the coronal voids show much weaker or no network structures. The detected flux imbalances fall in the range of imbalances found in QS areas of the same size.}
   {We conclude that coronal voids form because of locally reduced heating of the corona due to reduced magnetic flux density in the photosphere. This makes them a distinct class of (dark) structure, different from coronal holes.}


\maketitle

\section{Introduction}
Coronal holes \citep[CHs;][]{Waldmeister1956,Waldmeister1957} appear dark in  ultraviolet (UV) and X-ray observations of the solar corona. They are  cooler and less dense than the rest of the corona \citep{Munro1972,Cranmer2009}. During solar activity minima, large CHs form at the solar poles. These polar CHs disappear during solar activity maxima, while smaller CHs appear at lower latitudes. In the photosphere, CHs are harder to distinguish from the quiet Sun (QS) in terms of the magnetic flux density, as the absolute magnetic density in CHs is comparable to that of the QS. However, photospheric magnetic fields underlying CHs are distributed into one dominant magnetic polarity \citep[e.g.][]{Altschuler1972}. As a result of this, CHs are structured by open magnetic fields. Differently from the magnetically closed QS, this opens a channel for energy loss, namely the acceleration of gas that escapes the Sun’s gravitational field. This results in a reduced intensity in CHs at all coronal temperatures as compared to the QS.
    
In coronal observations in the extreme ultraviolet (EUV) around 174\,\AA\ by the high-resolution instrument of the Extreme Ultraviolet Imager \citep[EUI; see][]{EUI_instrument} on board Solar Orbiter \citep[][]{mueller:2020}, dark areas embedded in the QS can be identified. These coronal voids, as we call them, cover a wide range of scales, comparable in size to a few granules all the way up to a few supergranules (i.e. up to some 70\,Mm), making these features much smaller than typical CHs.
    
One possible interpretation of the voids could be that they are miniature CHs. Alternatively, the coronal voids could appear dark because they get less energy input from the magnetic structures at their base in the photosphere, and hence the plasma is less hot and less dense as compared to the surrounding QS. This could be the case if the magnetic field strength at their base is significantly lower than in the QS. \cite{Gonzalez2012} report `dead calm' areas with reduced magnetic activity, based on data obtained by the Imaging Magnetograph eXperiment \citep[IMaX;][]{Martinez_Pillet_2011} on the Sunrise observatory \citep{Solanki2010,Barthol2011TheSM,berkefeld2011wave,gandorfer2011filter} Hence, it is conceivable that the coronal voids are the coronal counterparts of such areas or generally of areas with reduced magnetic activity or magnetic flux density. 
        
To test both hypotheses, we studied the magnetic field in the photosphere underlying the coronal voids caught by the EUI instrument. We expected to find locally reduced magnetic activity, as in dead calm regions, and/or a net flux imbalance, as present in CHs. 
  
\section{Observations}
During Solar Orbiter's cruise phase on 2021 February 23 both the High Resolution Telescope \citep[HRT;][]{gandorfer18} of the Polarimetric and Helioseismic Imager \citep[SO/PHI;][]{PHI_instrument} and the High Resolution Imagers (HRI) of \eui\ observed a QS region at approximately the same time. Then Solar Orbiter was at a distance of 0.53\,AU to the Sun and the Solar-Orbiter--Sun--Earth angle measured {142\degree}. The telescopes were pointed at solar disk centre. According to the FITS file headers the image rotation of HRI EUV channel (\euv), the HRI Ly-$\alpha$ channel (\lya), and \hrt\ with respect to solar north were 0.3$\degree$, 0.5$\degree$, and 0.6$\degree$, respectively. In all images we show here north is approximately up and west is to the right. 
    
\subsection{SO/PHI observation}
\sophi\ is equipped with two telescopes, which can observe alternately. The Full Disc Telescope (FDT) maps the entire solar disk, while the \hrt\ has a smaller field of view (FOV) of $0.28\degree\times0.28\degree$. The spatial resolution of \sophi\ (and \eui, too) depends on the distance of Solar Orbiter from the Sun. 
    
\sophi\ scans the \ion{Fe}{i}\,6173\,\AA\ line at six different wavelengths and obtains a set of four polarisation states at each scan position. Hence, to obtain one set of data products, \sophi\ needs to record 24 images. After the subtraction of a dark and flat field correction and the demodulation of each image, the continuum intensity can be directly derived from the Stokes $I$ parameter. Assuming a Milne-Eddington atmosphere, the radiative transfer equation is inverted. This provides the magnetic field vector ${\bm B}=(B,\gamma,\phi)$ in the photosphere and the line-of-sight (LOS) velocity. The components $B$, $\gamma$, and $\phi$ of the magnetic field vector ${\bm B}$ denote the magnetic field strength, the magnetic field inclination relative to the LOS, and the magnetic field azimuth, respectively. 

The LOS magnetic field is defined as \blos$=B\cos\gamma$. The uncertainties in both $B$ and $\gamma$ are significant in quiet regions, like those studied in this paper. However, the uncertainties in these parameters are coupled in such a way that \blos\ suffers from much lower uncertainties, which generally makes \blos\ the most reliable measure of the magnetic field derived from polarimetric data in Zeeman split spectral lines. Further details are available in \citet{gandorfer18}, \citet{PHI_instrument}, and \citet{sinjan2022ground}.

\hrt\ took one full FOV observation of $2048\times2048$ pixels at the mid-observing time 17:00:45\,UTC (all times will be given in UTC) with a duration of 86~seconds. The \hrt\ has a plate scale of 0.5\arcsec. At the observing distance of 0.53\,AU this corresponds to 191\,km/pixel on the Sun. The LOS magnetogram we used in this study is a level 2 data product. The \blos\ map has additionally been corrected for geometrical distortion due to the instrument across the image. Because the observations were carried out near the centre of the solar disk, for the stronger, nearly vertical fields, \blos\ is very similar to the field strength. 

\subsection{EUI observations}    
Like \sophi, \eui\ is equipped with a Full Sun Imager (FSI) and with two high-resolution instruments. \hri{}'s EUV channel, \euv, is centred around 174\,\AA\ including a prominent \ion{Fe}{x} line that samples gas at approximately 1\,MK. The other channel, \lya, records the solar scene in the Ly-$\alpha$ line of hydrogen at 1216\,\AA\ originating from the chromosphere. Since the \hri\ channels employ separate telescopes, simultaneous observations are possible. The \hri\ detectors record images with a size of up to $2048\times2048$~pixels with a FOV comparable to \hrt. 
    
Due to technical issues the two channels of \hri\ did not observe in parallel. For the Ly-$\alpha$ channel a time series is available from 16:55:15 to 16:58:55 with a cadence of 5~seconds. The 174\,\AA\ channel started observing at 17:13:25. This time series with a 2-second cadence ended at 17:20:59. For our analysis, we chose the \lya\ and \euv\ observations closest to the selected \hrt\ observation, namely at 16:58:55 and 17:13:25, respectively.
\euv\ data have a plate scale of 0.492\arcsec, which corresponds to 188\,km/pixel at the solar distance of 0.53\,AU. The \lya\ data were binned by a factor of four and have a plate scale of 1.028\arcsec (392\,km/pixel at the given solar distance). The \eui\ observations used in this study are level~2 data products.\footnote{\url{https://doi.org/10.24414/2qfw-tr95}}
\citet{Kahil2022} also used this \euv\ observation as well as the \sophi\ \blos\ map.

\subsection{Alignment of PHI and EUI}    
Due to temperature changes within the spacecraft and instruments, the alignment between instruments and channels is expected to change over time. The EUV images therefore need to be aligned with the magnetogram used here. Because the magnetic network is apparent both in \blos\ and in \lya\ an alignment in sequence (\blos\ to \lya\ to \euv) yields the best result. We followed the procedure for the alignment described by \citet{Kahil2022}, but we additionally adjusted the plate scale of the \blos\ and \lya\ images by rescaling them to the \euv\ plate scale of 0.492\arcsec (188\,km/pixel on the Sun at 0.53\,AU). The \blos\ map is shifted southwards with respect to the \euv\ observation since the overlap in FOV of both instruments is not exact. Consequently, there is no magnetic field measurement available for the upper part (ca. 10\%) of the \euv\ image. The \lya\ observation was noticed to be slightly shifted westwards during the alignment procedure. 

\section{Results}
We first defined the coronal voids via an intensity threshold and subsequently tested both hypotheses for the formation of coronal voids by analysing the photospheric magnetic field. 
    
\begin{table*}
\centering
    \caption[]{Properties of major coronal voids and QS.}
    \label{Table_coronal_voids}  
        \begin{tabular}{c c c c c c}          
        \hline\hline                       
        Coronal void\tablefootmark{a} & Size\tablefootmark{c} & Avg. coronal & $\langle$|\blos|$\rangle$\tablefootmark{e} & \multicolumn{2}{c}{Imbalance in magnetic flux\tablefootmark{f}}\\
        & [Mm] & intensity\tablefootmark{d} [DN/s] & [G] & relative $\delta F$ [\%] & absolute $\Delta F$ [G]\\    
        \hline                                 
    1 & 77 & 735 & 12.1 & 12.2 & $+$1.5 \\      
    2 & 21 & 756 & 10.1 & 11.3 & $+$1.2 \\
    3 & 49 & 708 & 11.2 & 16.1 & $-$1.8 \\
    4 & 26 & 739 & 9.8 & 26.1 & $-$2.6 \\
    5 & 21 & 773 & 12.2 & 42.5 & $-$5.2 \\
    6 & 60 & 741 & 12.0 & 8.2 & $-$1.0 \\
    \hline  
    Surrounding QS\tablefootmark{b} && 1122 & 15.9 & 0.9 & 0.1 \\
    Whole FOV && 1094 &15.7 & 1.0 & 0.2 \\
    \hline
\end{tabular}
\tablefoot{%
\tablefoottext{a}{The outlines of the major coronal voids are shown in Fig.~\ref{EUI+PHI_overview}.}
\tablefoottext{b}{The surrounding QS is the whole FOV displayed in Fig.~\ref{EUI+PHI_overview} without the major coronal voids.}
\tablefoottext{c}{The size is the diameter of a circle covering the same area as the void (see Sect.~\ref{S:define.voids}).}
\tablefoottext{d}{Average intensity seen by \euv\ in the 174\,\AA\ channel of EUI.}
\tablefoottext{e}{For the definition of |\blos|, see Sect.~\ref{S:B.in.voids}.}
\tablefoottext{f}{Flux imbalances as defined in Eqs.~\ref{Eq.rel.imbalance} and \ref{Eq.abs.imbalance} (see Sect.~\ref{subsection_flux_imbalance}).}
}
\end{table*}

\begin{figure*}
\centering
     \includegraphics[width=\textwidth]{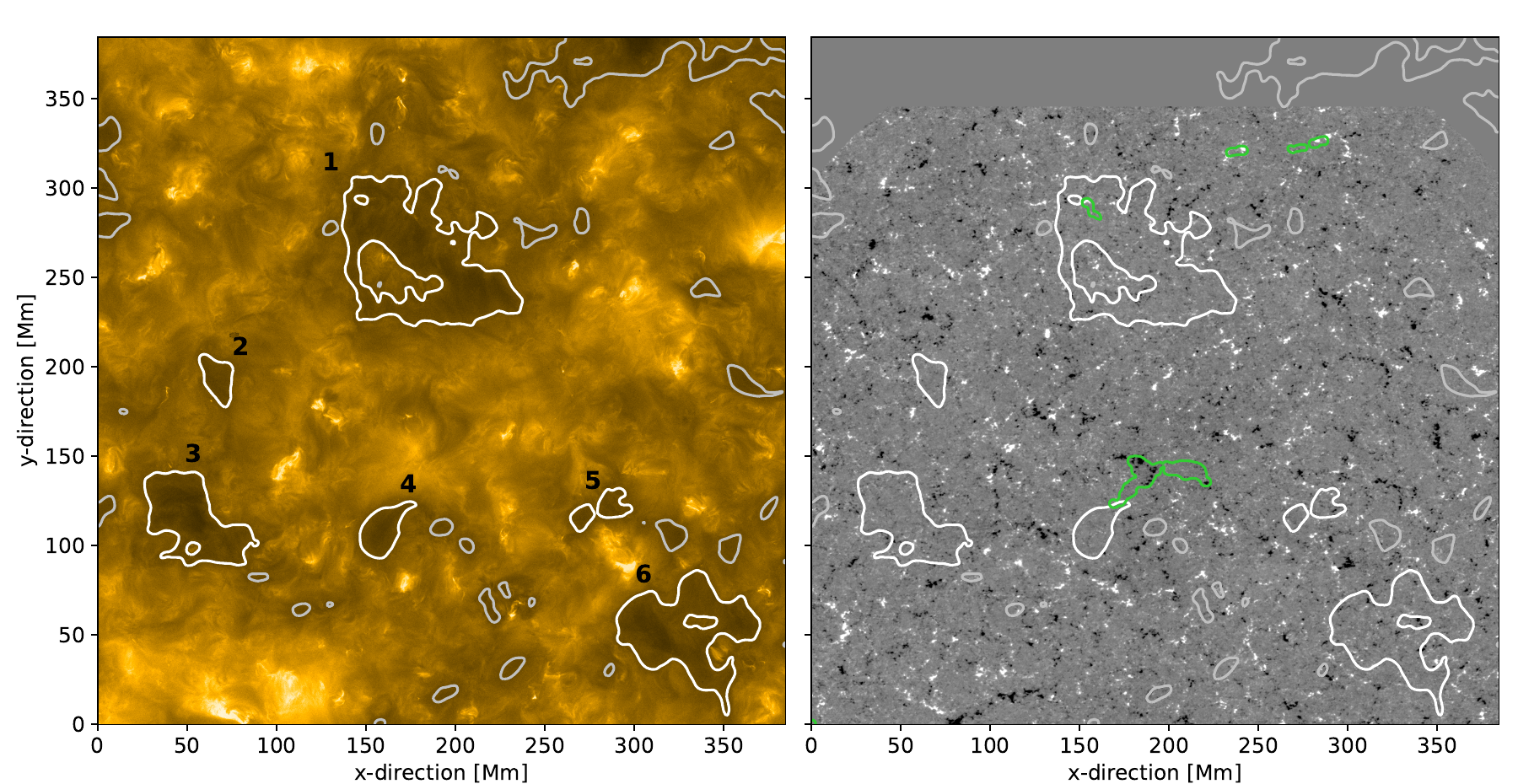}
     \caption{Overview of observations. 
     These high-resolution images show the (almost) full FOV of EUI \euv\ (left) and \hrt\ (right) and were taken on 2021 February 23 at 17:13:26 and 17:00:45, respectively. Disk centre (as seen from Solar Orbiter) is roughly in the centre of the images, and north is up. Coordinates are given in Mm on the Sun. The \euv\ image around 174\,\AA\ is shown on a linear scale, and the \hrt\ map shows the LOS magnetogram saturated at $\pm$50\,G with white and black indicating positive and negative polarities, respectively. Because the FOV of PHI is shifted somewhat towards the south with respect to EUI, the top part of the magnetogram does not show data (flat grey). All white or grey contours show EUI intensity levels of 75\% of the average intensity and highlight the dark coronal voids seen in the \euv. The six major voids are indicated by thick white outlines and numbered according to the list in Table \ref{Table_coronal_voids} (see Sect. \ref{S:define.voids}). The green contours mark the regions on the surface from where the open field lines originate. For the discussion, see Sect. \ref{subsection_extrapolation}.}
     \label{EUI+PHI_overview}
\end{figure*}

\subsection{Defining coronal voids}\label{S:define.voids}
The coronal voids are easily picked out by eye in the \euv\ images (Fig.\ \ref{EUI+PHI_overview}, left panel). To capture their outline (i.e.\ shape and size), we used an iso-contour of the intensity in a smoothed image. We smoothed the image to avoid the outline appearing overly ragged. For this we convolved the \euv\ image with a Gaussian kernel with a full width at half maximum (FWHM) of about 4\,Mm that corresponds to the size of the network patches visible in the QS EUV data. We then used a threshold of 75\% of the mean intensity in the FOV. This captures the outline of the dark voids quite well, as visible in Fig.\ \ref{EUI+PHI_overview} (left panel) where we overplot this contour (based on the smoothed image) on top of the original \euv\ image. A coronal void is then the region enclosed by this intensity contour. We experimented with other values for the threshold, but found that the results did not change significantly, even though, naturally, the area of a void depends somewhat on the threshold (larger voids are found if the threshold is increased).

The coronal voids cover about 10\% of the full FOV and exhibit a mean intensity of the order of 60 to 70\% of the QS level. Table~\ref{Table_coronal_voids} lists details on the coronal voids identified (and numbered) in Fig.\ \ref{EUI+PHI_overview} and on the QS. 
    
The coronal voids display a variety of sizes and shapes. For an easy comparison with, for example, the supergranular scale, we defined the linear dimension of a coronal void as the diameter of a circle that would cover the same area as the coronal void. 
For our study we concentrated on the larger coronal voids. We refer to these voids with a size of at least a supergranule (ca. 20\,Mm) as major voids (Table~\ref{Table_coronal_voids}), and identified five structures meeting this criterion (voids no. 1 to 4 and 6 as marked in Fig.~\ref{EUI+PHI_overview}). In addition, we considered a further void to be major (no. 5), which consists of two smaller voids very close to each other, almost touching (combined, the two parts have a size of just above 20\,Mm). We did not consider a large void partially visible in the upper right because it is at the edge of the EUI FOV (and would not be covered by \hrt\ at all).
    
Hence, we have six fully distinguishable major coronal voids whose magnetic properties we can study. Smaller dark structures that are below the supergranular scale are not considered further, as they may simply be a reflection of dark inter-network areas (small grey contours in Fig.\ \ref{EUI+PHI_overview}).

In order to put the threshold used for the definition of coronal voids into context, we compared it to thresholds employed in studies of polar or equatorial CHs.
For the detection of CHs, different values of the detection threshold are used, depending on the temperature of the plasma imaged by the respective observation. Currently, often the 193\,\AA\ passband of the Atmospheric Imaging Assembly \citep[AIA;][]{SDO/AIA} on board the Solar Dynamics Observatory \citep[SDO;][]{Pesnell2012} is used to characterise the size and shape of CHs. The 193\,\AA\ band includes \ion{Fe}{xii} lines and has its peak response at around 1.6\,MK \citep{SDO/AIA}. For this channel often an intensity threshold of 35\% to 40\% of the median solar disk intensity is applied \citep[e.g.][]{Hofmeister_2017,Heinemann2019}. This contrast is reduced in the emission below 1\,MK that is sampled by the AIA 171\,\AA\ channel \citep{Garton2018_AIA171inCH}, that is, a larger threshold has to be applied when outlining a polar or equatorial CH.
So overall, we consider the threshold we used for the coronal voids larger, but consistent with, the definition of CHs.

\subsection{Chromospheric counterparts of coronal voids}\label{section_on_lyman-alpha}

\begin{figure}
\centering
    \resizebox{\hsize}{!}{\includegraphics{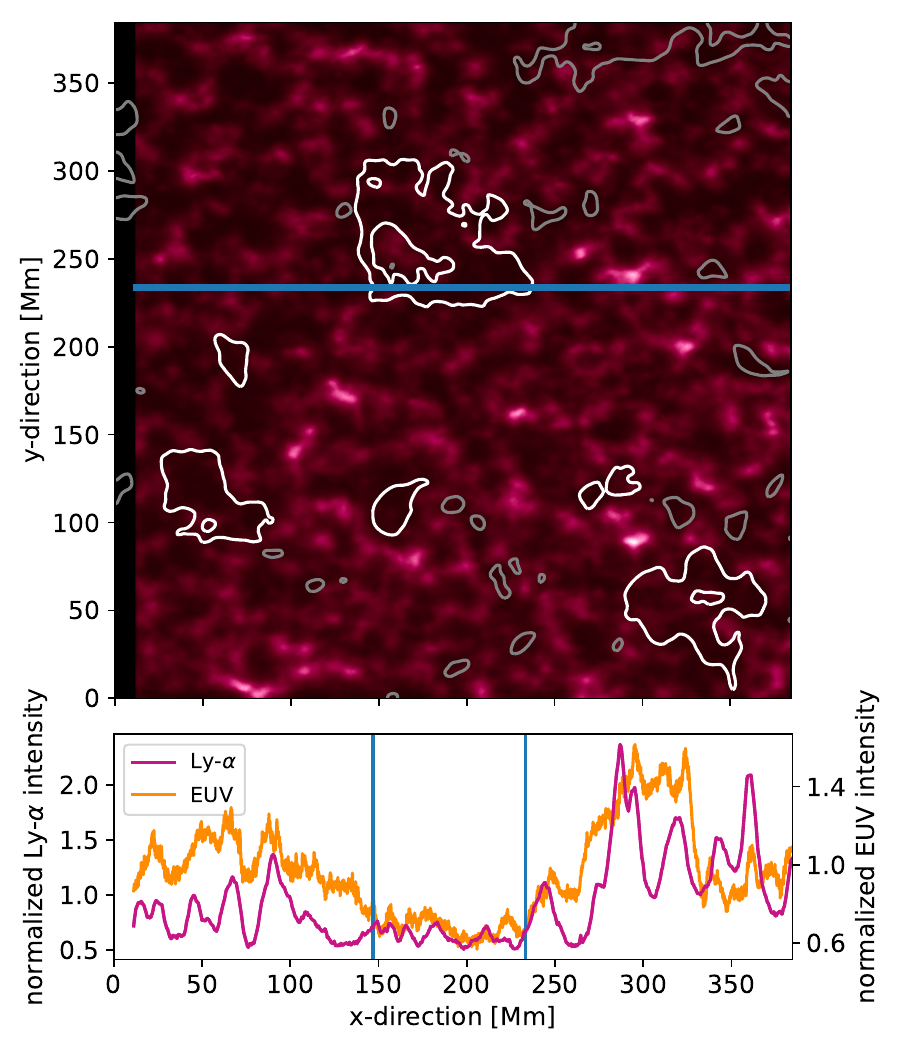}}
     \caption{Connection of coronal voids to the chromosphere. 
     The upper panel shows an intensity image of the region of interest as recorded by \lya\ on 2021 February 23 at 16:58:55, with the same FOV and within about a quarter-hour of the snapshots shown in Fig.~\ref{EUI+PHI_overview}. The contour lines are the same as in Fig.~\ref{EUI+PHI_overview}, outlining the coronal voids.
     The lower panel shows a cut through the image along the blue line in the Ly-$\alpha$ image. Plotted is the intensity in the \lya\ (red) and \euv\ (orange) channels. To reduce the noise, the cuts have been averaged in the $y$-direction over 10 pixels. The dark coronal void is seen in the cut in Ly-$\alpha$ as well as in the EUV at 174\,\AA\ (vertical blue lines indicate the extent of the void intersected by the cut).
     See Sect.\ \ref{section_on_lyman-alpha} for details.}
     \label{Lyman_alpha}
\end{figure}

The Ly-$\alpha$ channel of \hri\ enables us to look for chromospheric counterparts of coronal voids.
The upper panel in Fig.~\ref{Lyman_alpha} shows the \lya\ image with the boundaries of the voids identified in the \euv\ channel overplotted. In general, the Ly-$\alpha$ channel shows reduced intensity where we see coronal voids in the 174\,\AA\ channel. In the \lya\ image inside the coronal voids only very faint structures are visible.
To quantify the reduced emission in the voids of both \lya\ and \euv\ we show a cut along the $x$-direction through the largest of the coronal voids (red and orange curves, respectively, in the lower panel of Fig.~\ref{Lyman_alpha}). To reduce noise, we averaged the \lya\ and \euv\ intensities over 10 pixels in the $y$-direction before plotting. We also see similar behaviours for the other major voids.

In the void, the Ly-$\alpha$ emission is lower than in its surroundings. Small brightness fluctuations can be seen in the void in Ly-$\alpha$, but these are much weaker than the large fluctuations in the QS. Outside the void, the brightest features in Ly-$\alpha$ are related to the magnetic network, in line with the expectation that the magnetic field and chromospheric emission in Ly-$\alpha$ are correlated. In the voids, however, we see little evidence for the presence of stronger patches of the magnetic network (see Sect.\ \ref{S:mag.field}) and thus we consider it unlikely that these patches of very slightly enhanced Ly-$\alpha$ emission would be connected to the magnetic network.

\subsection{Photospheric magnetic field} \label{S:mag.field}
In order to determine whether the reduced EUV intensity inside the voids might be due to reduced heating related to weaker photospheric magnetic fields, we considered the {\hrt} \blos\ map.
    
\subsubsection{Magnetic field in coronal voids and in the QS}\label{S:B.in.voids}

\begin{figure}
\centering
     \resizebox{\hsize}{!}{\includegraphics{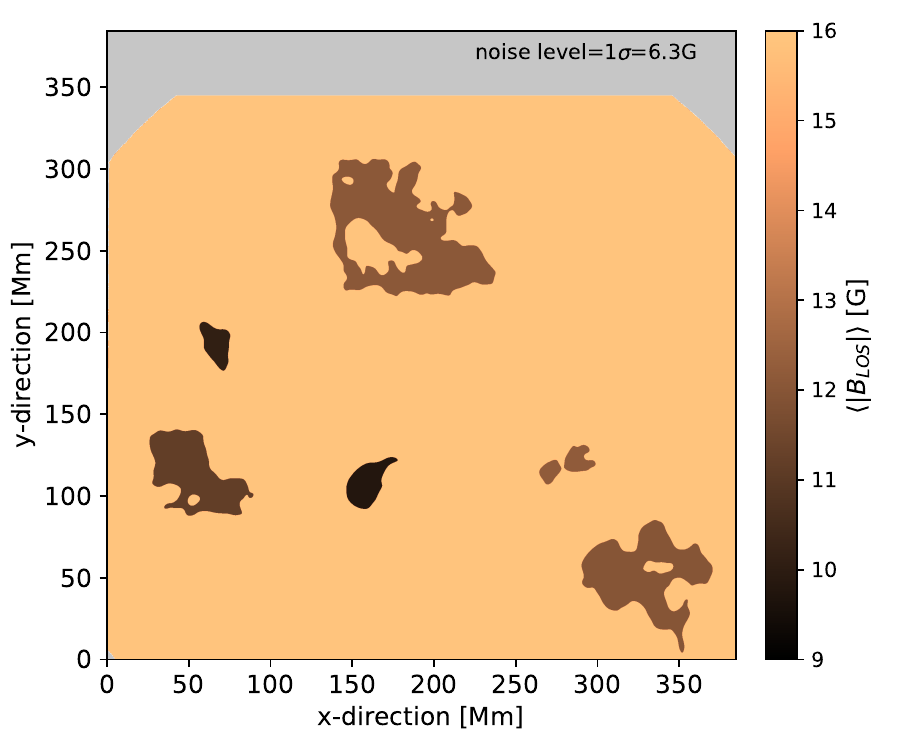}}
     \caption{Illustration of the absolute LOS magnetic field strength in major voids and QS.
     For the six major voids, the average of the LOS field, |\blos|, is calculated and then plotted as a single value within the contour of the respective void (see also Fig.~\ref{EUI+PHI_overview} and Table~\ref{Table_coronal_voids}). Likewise, the QS region is coloured corresponding to the average |\blos| of the QS.
     See Sect.~\ref{S:B.in.voids}.}
     \label{mean_abs_blos}
\end{figure}

For each of the major voids and for the surrounding QS the mean of the unsigned LOS magnetic field was determined. We obtained this value by first projecting the contours of the coronal voids onto the \blos\ map, then computing the absolute value of \blos\ wherever it is above the noise level of $1\sigma=6.3$\,G \citep{sinjan2022ground}. Finally, we took the average of these values over a whole selected area (an individual void or the QS).
This is illustrated in Fig.~\ref{mean_abs_blos}, where the respective mean value has been displayed in the entire area covered by the respective coronal void or the QS. The exact values are listed in Table~\ref{Table_coronal_voids}. 

The mean |\blos| inside the voids ranges between 9.8 and 12.2\,G, while it is 15.9\,G in the QS; in other words, the field is weaker by a factor of 1.3 to 1.6 in the voids compared to the QS. Therefore, on average the coronal voids show a clearly weaker photospheric magnetic flux than the surrounding QS.

\subsubsection{Strong magnetic fields avoid coronal voids}\label{S:strong.B.in.voids}

\begin{figure*} [h]
\centering
     \includegraphics[width=\textwidth]{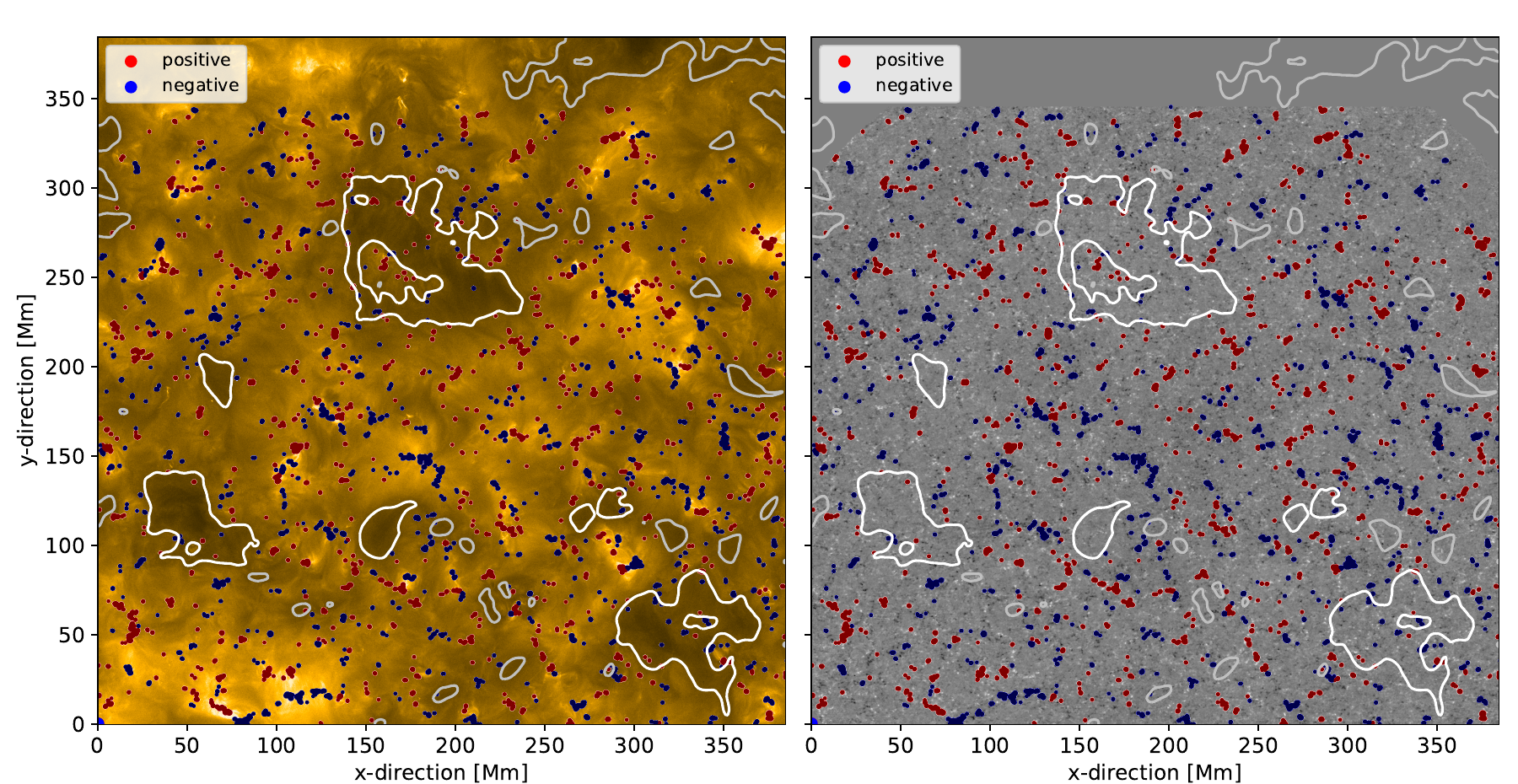}
     \caption{Strong-field regions in voids and QS. 
     The background image is the same as in Fig.~\ref{EUI+PHI_overview}, i.e. the 174\,\AA\ image from \euv\ (left) and the magnetogram from \hrt\ (right).
     To highlight strong-field regions, we colour the regions of magnetic field with a strength above $\pm$50\,G in red and blue for positive and negative magnetic polarities, respectively.
     See Sect.~\ref{S:strong.B.in.voids}.
     }
     \label{location_strong_pixels}
\end{figure*}

\begin{figure}
\centering
    \resizebox{\hsize}{!}{\includegraphics{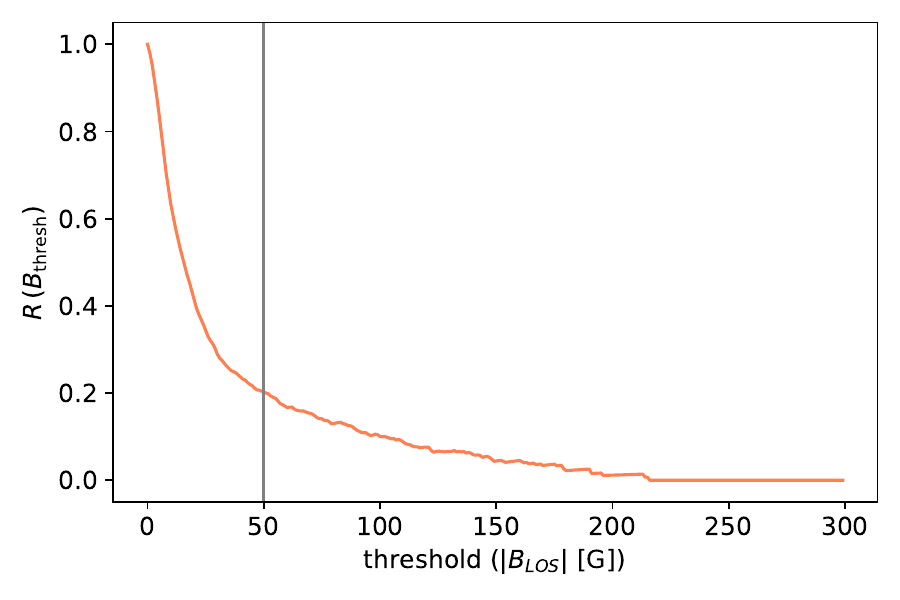}}
     \caption{Ratio, $R$, of areas of voids to QS with |\blos| above a given threshold plotted vs this threshold in |\blos|. The area ratio is normalised such that it reaches unity as |\blos| goes to zero (see Eq.~\ref{Eq:ratio.B.thresh}). The vertical line indicates the 50\,G pixel distribution shown in Fig.~\ref{location_strong_pixels}.}
     \label{ratio_pixel_distribution}
\end{figure}

The occurrence of pixels with stronger |\blos| is quite different inside the coronal voids compared to the surrounding QS. Fig.~\ref{location_strong_pixels} illustrates this, where we mask the locations of strong fields with $|B_\mathrm{{LOS}}| > 50$\,G for both magnetic polarities in the \euv\ and the \blos\ map.
There are clearly fewer pixels with stronger \blos\ per area inside the voids than outside. In fact, very few strong field regions can be found inside the voids. 
In addition, those few strong magnetic patches in the void are on average smaller than in the surrounding QS. Therefore, the magnetic network is either particularly weak or entirely absent within the voids.

As a next step, we quantified the different coverage of magnetic fields of different strength in the voids as compared to the QS.
For this, we calculated the number of pixels above a given threshold $B_{\rm{thresh}}$ of the absolute LOS magnetic field strength, |\blos|, and we did this separately for all six major voids together and for the QS without these major voids. Essentially, the ratio of these values is the ratio of area covered by fields above $B_{\rm{thresh}}$ in voids compared to the QS. We normalised this by the ratio for $B_{\rm{thresh}}{=}0$\,G, which we call $R_0$ and which is essentially the area ratio of the voids to the QS outside the voids. This then yields the normalised area ratio for fields above $B_{\rm{thresh}}$,
\begin{equation}\label{Eq:ratio.B.thresh}
R\,({B_{\rm{thresh}}}) ~~=~~ \frac{1}{R_0} ~ 
\frac{~\sum \mbox{pixels in voids}\,(|B_{\rm{LOS}}|>B_{\rm{thresh}})~}
    {\sum \mbox{pixels in QS}\,(|B_{\rm{LOS}}|>B_{\rm{thresh}})} ~.
\end{equation}
We calculated this ratio for values of $B_{\rm{thresh}}$ from 0 to 300\,G and display the result in Fig.~\ref{ratio_pixel_distribution}.

We see a quick drop of the area covered in voids with the threshold field $B_{\rm{thresh}}$. At $B_{\rm{thresh}}{=}50$\,G the ratio dropped already to about 1/5 (see the vertical line in Fig.~\ref{ratio_pixel_distribution}). This implies that when comparing a region of a void and of QS covering the same area, in the QS there would be 5 times more pixels with |\blos| above 50\,G. Just above 200\,G the ratio $R$ drops to zero indicating that there are no more pixels with |\blos| that strong or stronger inside the coronal voids, while pixels with stronger |\blos| can still be found in the QS.
This underlines that within coronal voids the magnetic field is not only weak on average, but that there are very few regions with stronger LOS field strength.
     
\begin{figure*} [h]
\centering
     \includegraphics[width=\textwidth]{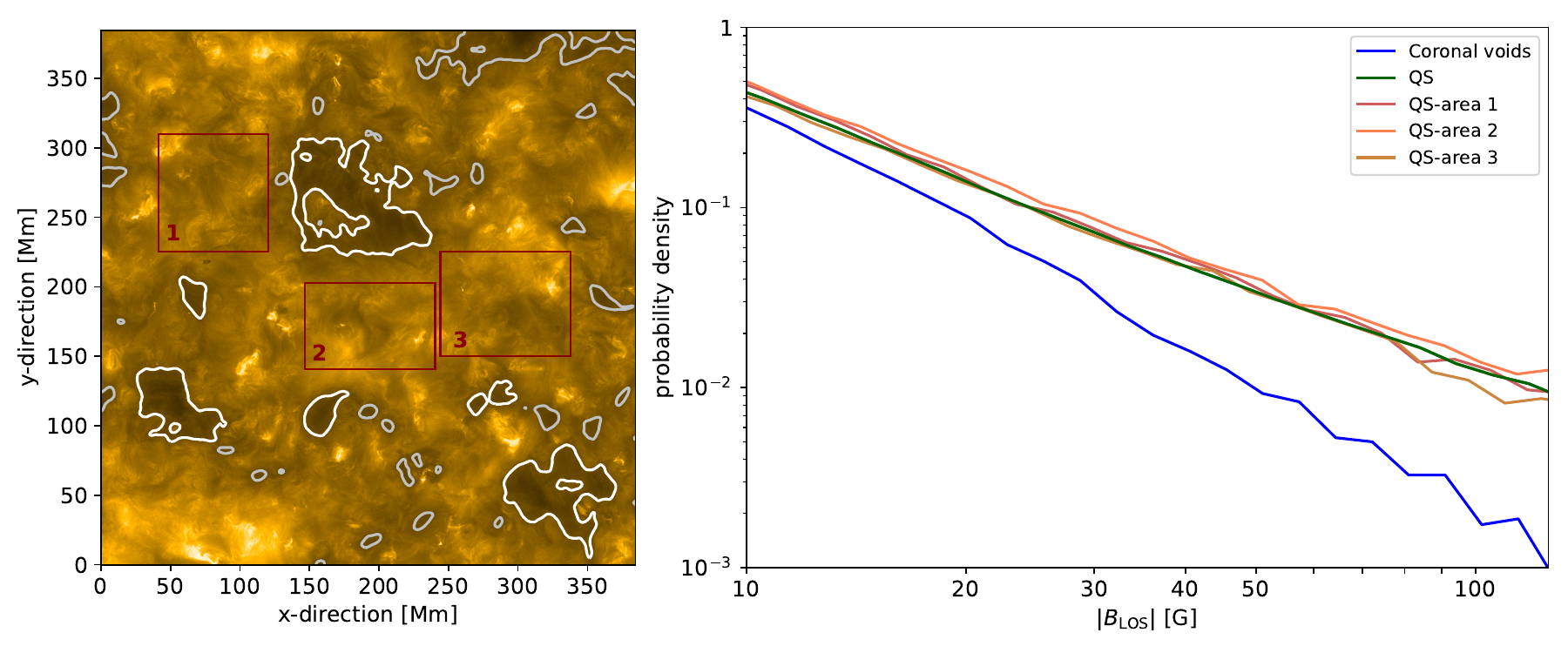}
     \caption{Histograms of the absolute LOS magnetic field in coronal voids and the QS.
     The left panel shows the \euv\ image, similar to Fig.~\ref{EUI+PHI_overview}, along with the regions of interest, from which we derived the histograms of |\blos| that are displayed in the right panel.
     The solid blue line shows the histogram for all major voids combined (thick white contours in the left panel). The red, orange, and brown lines show the histograms within the three squares in the left panel, all covering small parts of QS. The green histogram is for the whole FOV except for the major voids.
     See Sect.~\ref{S:strong.B.in.voids}.}
     \label{doublelog_absblos_hist}
\end{figure*}

The different distribution of stronger magnetic fields in coronal voids and the QS also becomes apparent in a double-logarithmic histogram of |\blos| (using bins equally spaced in log(|\blos|) with a width of 0.05; shown in Fig.~\ref{doublelog_absblos_hist}, right panel). We created a histogram of |\blos| for all the coronal voids together (i.e. all voids are represented by a single histogram) and a separate one for the surrounding QS (i.e. the full FOV except the voids). We see a power-law-like histogram for voids and QS, with the slope of the voids histogram being much steeper. Hence, inside the voids the number of strong-field patches drops much faster than in the QS, illustrating that the spatial density of strong field patches in the voids is very low as compared to the QS.
    
Given that the QS area is considerably larger than the area of the voids, there may be smaller regions within the QS with an equally low density of strong-field patches as the voids. To test this, we also investigated the histograms for three QS regions of a size comparable to the largest void. We chose these regions more or less randomly as rectangles 1 to 3 in the left panel of Fig.~\ref{doublelog_absblos_hist}. These QS regions show a power-law distribution essentially identical to the histogram of the whole QS area. This underlines that the difference between the histograms of the voids and the QS is real and is not skewed by the different areas of the voids and the QS.

\begin{figure}
\centering
    \resizebox{\hsize}{!}{\includegraphics{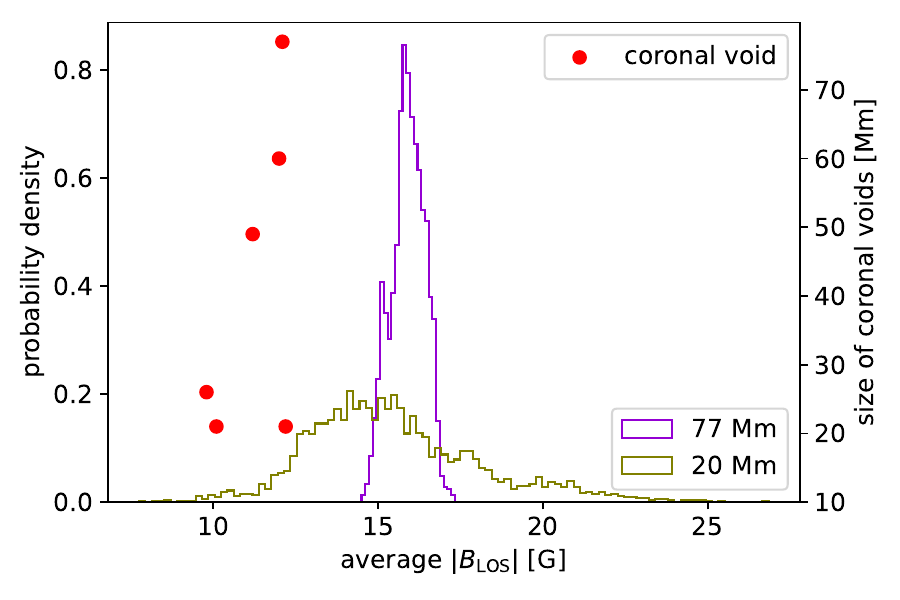}}
    \caption{Average |\blos| in QS areas. The QS areas of the whole FOV are sub-divided into identical tiles. For each tile the mean |\blos| is calculated. Histograms of the mean |\blos| for tiles with a size corresponding to 77 and 20\,Mm are shown in purple and olive, respectively. The red dots mark the mean |\blos| values derived for the six major voids (cf. Table~\ref{Table_coronal_voids}) as a function of their size (right axis). See Sect.~\ref{S:strong.B.in.voids}.}
    \label{QS_histograms}
\end{figure}

We further tested the distribution of the mean |\blos| in QS areas. We first tiled the QS areas of the whole FOV into small squares, each roughly the size of a void. We then calculated the mean |\blos| above the noise level of $1\sigma$ inside each tile. To enhance statistics (i.e. the number of tiles), we allowed the tiles to overlap (by introducing a shift of 5\,Mm between two tiles). We carried out this procedure for two tile sizes, one with an area corresponding to the smallest major void (20\,Mm diameter, corresponding to a tile with $\sqrt{\pi}/2\,{\times}\,20$\,Mm side length) and another one corresponding to the 77\,Mm sized void. This yields two histograms for the mean |\blos|, which we compared to the values derived for the major coronal voids (Fig.~\ref{QS_histograms}). The values for the coronal voids clearly differ from the QS histograms. All of them exhibit weaker |\blos| values as expected from the QS histograms and are located at the lowest end of the QS distributions. This further supports the claim that strong magnetic fields avoid coronal voids.

\subsubsection{Weak magnetic fields define coronal voids}\label{S:weak.B.def.voids}

\begin{figure}
\centering
    \resizebox{\hsize}{!}{\includegraphics{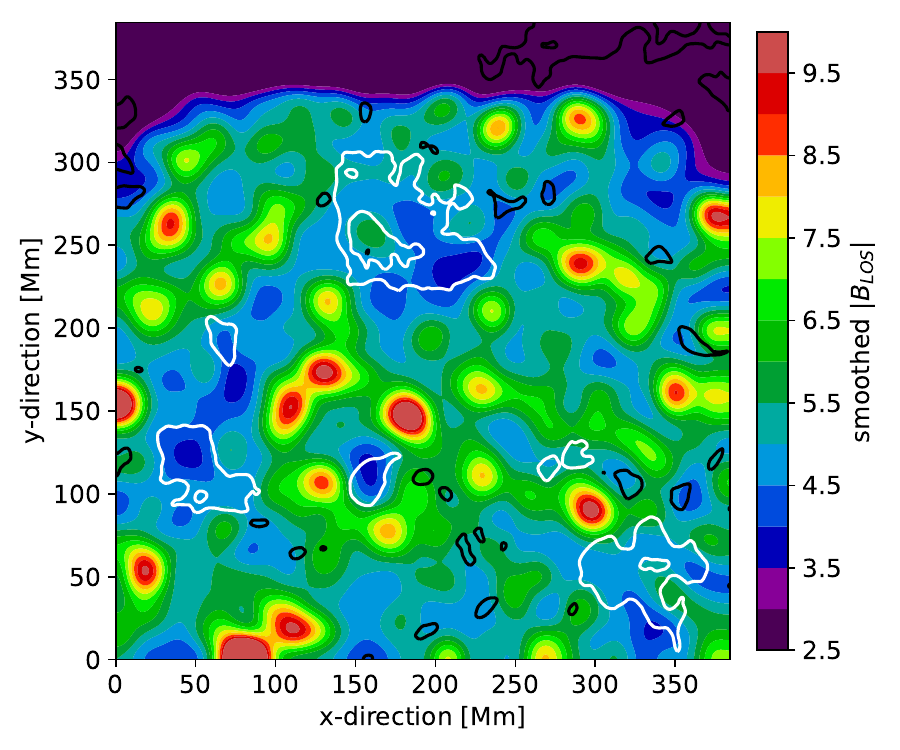}}
    \caption{Weak magnetic fields and coronal voids.
    To highlight the regions of weaker magnetic field, we smoothed the LOS magnetic field strength |\blos| using a Gaussian kernel of 20\,Mm FWHM. To emphasise the structuring, we use a discrete colour table. The weak-field regions are found in areas displayed mainly in blue. The contours of coronal voids are overplotted (the six major voids listed in Table~\ref{Table_coronal_voids} in thick white, smaller voids in black; same contours as in Fig.~\ref{EUI+PHI_overview}).
    See Sect.~\ref{S:weak.B.def.voids}.}
    \label{smoothedBLOS}
\end{figure}

After having demonstrated in Sect.~\ref{S:strong.B.in.voids} that coronal voids are areas of weak magnetic activity, we now address the question whether all regions of low magnetic field harbour coronal voids. 
The QS magnetic field mostly forms a small-scale salt \& pepper pattern. We therefore considered the unsigned LOS magnetic field, |\blos|, and reduced the noisiness of the signal by spatially averaging the |\blos| map. Because the major voids we were looking for have sizes of 20\,Mm or more (cf. Table \ref{Table_coronal_voids}), we smoothed the |\blos| map using a Gaussian kernel with a FWHM of approximately 20\,Mm (115 pixels). The result is shown in Fig.~\ref{smoothedBLOS}.
Of course, because of the smoothing the values in this map are lower than the values for |\blos| in the original maps used, for example. in 
Figs.~\ref{ratio_pixel_distribution} or \ref{doublelog_absblos_hist}.

The smoothed |\blos| map shows that many of the weak-field regions are covered by coronal voids.
In Fig.~\ref{smoothedBLOS} the larger patches of weak field (in dark to light blue) are covered by the major voids. The additional smaller weak-field patches are mostly covered by smaller coronal voids (with sizes below 20\,Mm; black contours).

There are some areas that exhibit weak magnetic field strength and at the same time do not harbour voids (either small or large). An example is found near $x{=}80$\,Mm and $y{=}170$\,Mm, another one at the bottom-left of the FOV shown in Fig.~\ref{smoothedBLOS}.
In cases like this, it seems that coronal structures arch above the weak-field regions connecting patches of opposite magnetic polarity on either side of the weak-field region. Hence, in those weak-field regions we do not detect a small coronal void because QS coronal loops are crossing them (e.g. in the bottom-left part of the FOV).

From this we conclude that coronal voids are found above many areas of low average magnetic field strength in the photosphere. In most of the exceptions where there is no coronal void above a weak-field region, larger overarching coronal loop structures are seen that connect stronger magnetic patches on either side of the weak-field region.

\subsection{Flux imbalance} \label{subsection_flux_imbalance}

\begin{figure}
\centering
    \resizebox{\hsize}{!}{\includegraphics{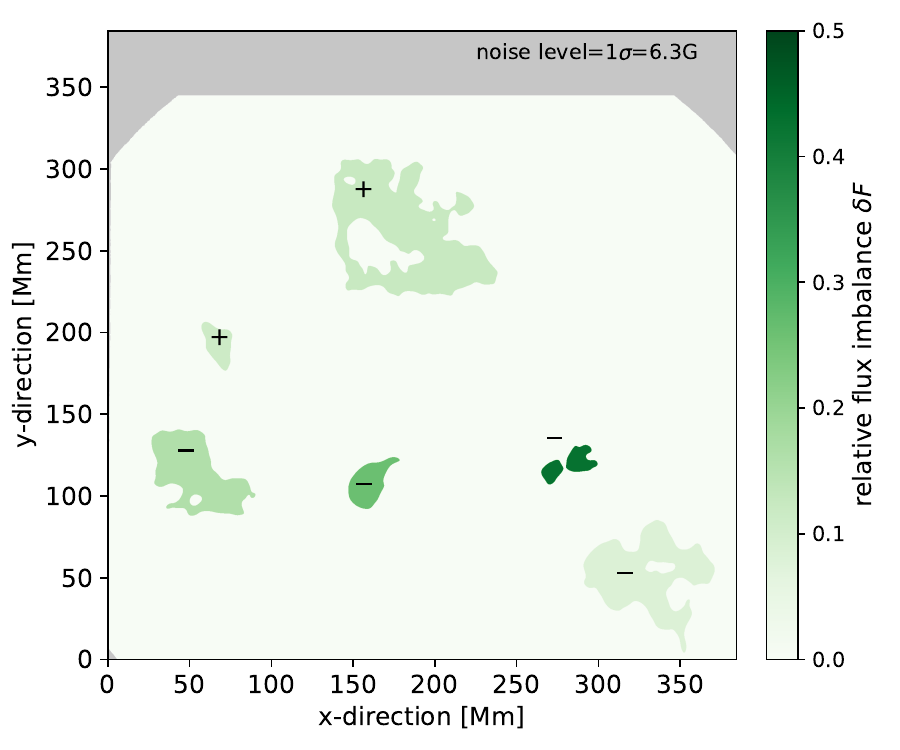}}
    \caption{Illustration of the relative magnetic flux imbalance in the major voids and the QS. 
     For each of the six major voids, we calculated the relative flux imbalance as defined in Eq.~\ref{Eq.rel.imbalance}. The respective imbalances are then plotted as a single value within the contour of the respective void (see also Fig.~\ref{EUI+PHI_overview} and Table~\ref{Table_coronal_voids}). The $+$ and $-$ signs overplotted on or just beside the voids indicate the net magnetic polarity within each void. Likewise, the QS region is coloured corresponding to the (negligible) flux imbalance of the QS.
     See Sect.~\ref{subsection_flux_imbalance}}
     \label{imbalance_dF}
\end{figure}

\begin{figure}
\centering
    \resizebox{\hsize}{!}{\includegraphics{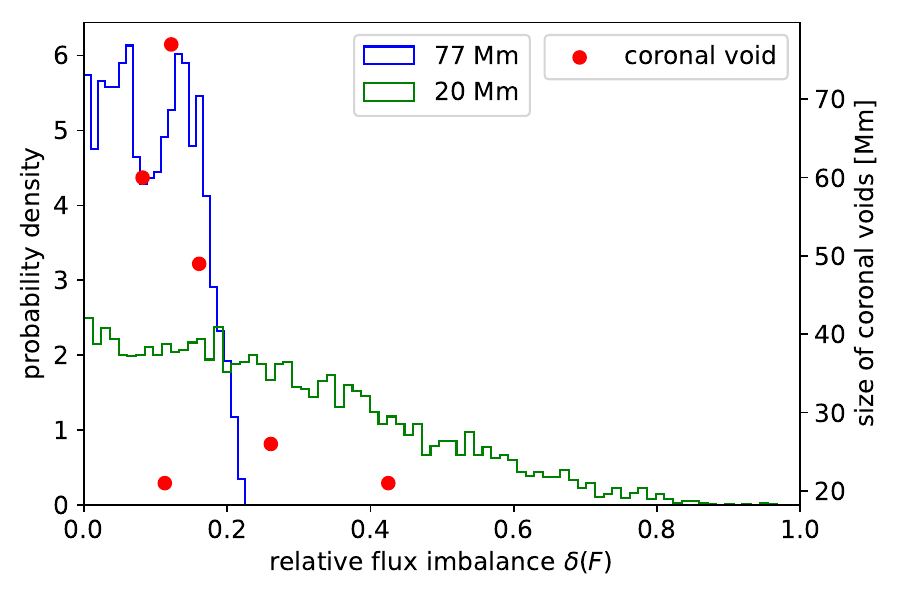}}
    \caption{ Relative flux imbalance in QS areas.
    The QS areas of the whole FOV are sub-divided into identical tiles. In each tile the relative flux imbalance is calculated according to Eq.~\ref{Eq.rel.imbalance}. The histograms of imbalances for tiles with a size corresponding to 77 and 20 Mm are shown in blue and green, respectively.
    The red dots depict the flux imbalance in the six major voids (cf.\ Table~\ref{Table_coronal_voids}) as a function of their size (right axis).
    See Sect.~\ref{subsection_flux_imbalance}.}
    \label{distribution_imbalance_in_QS}
\end{figure}

To test the hypothesis whether a coronal void could also be a miniature version of a CH, we checked the level of flux imbalance inside each coronal void and compared these values to actual CHs. The presence of a flux imbalance is a necessary condition for this analogy. Of course, a second step would then be to test if the voids are magnetically open or close back down to the solar surface into another void or QS patches.
    
We defined the relative imbalance in the magnetic flux as the absolute value of the average \blos\ for a given area, divided by the average of the absolute value of \blos,
\begin{equation}\label{Eq.rel.imbalance}
\delta F = {|\left< B_{\rm LOS,t}\right> | \over \left< |B_{\rm LOS,t}| \right>} \,,
\end{equation}
where quantities between brackets $\left< ... \right>$ are spatial averages and $B_{\rm LOS,t}$ is the \blos\ value above some threshold. Here we set the threshold to the noise level \cite[6.3\,G;][]{sinjan2022ground} to reduce spurious effect due to noise. 
We computed the relative imbalance, $\delta F$, separately for each major coronal void and for the QS (without major voids). This is illustrated in Fig.~\ref{imbalance_dF} and the respective values are listed in Table \ref{Table_coronal_voids}.
The relative flux imbalance in the voids is in the range from about 10\% to 25\%, with the only exception being the smallest void (number 5) showing about 40\% (Table \ref{Table_coronal_voids}). Compared to this the QS imbalance is negligible (1\% or less).
While the flux imbalances for the major voids might seem significant, they are still much smaller than the corresponding values found in low-latitude CHs, where \cite{Wiegelmann2004} and \cite{Hofmeister_2017} found net flux imbalances of $77\%\pm14$\% and $49\%\pm16$\%, respectively.
    
To better judge the significance of the relative flux imbalance, we also checked the absolute imbalance in the magnetic flux, that is, the mean \blos\ of a given area above the noise threshold, 
\begin{equation}\label{Eq.abs.imbalance}
\Delta F = \left<B_{\rm LOS,t}\right> .
\end{equation}
The resulting values for the major voids and the QS are listed in Table \ref{Table_coronal_voids}, too.
Unsurprisingly, for the QS observation covering an area of nearly $400\times400$\,Mm, the magnetic flux in the full FOV is almost balanced. The major voids, on the other hand, show absolute values of $\Delta F$ between about 1 and 2\,G, again with only void 5 showing a larger value (of about 5\,G). Just as for the relative imbalance, these values are significantly smaller than the corresponding values found in CHs \cite[e.g. 7.6\,G reported by][]{Wiegelmann2004}.
    
The magnetic polarity (i.e. the sign of $\Delta F$; cf Table \ref{Table_coronal_voids}) is not the same for the major voids. Hence, we cannot consider that the magnetic field from these voids might join higher up in the corona, where it then, together, could be connected to the solar wind.
    
The flux imbalances we see in the major voids could also be an artefact of their (small) size and the correspondingly poor statistics of the magnetic patches. The whole QS region (without the major voids) is essentially flux balanced, but it is also more than a factor of 10 larger in area than all the major voids together. To check the significance of the flux imbalance in the major voids, we had to do a  test that checks the flux imbalance of the QS in smaller regions that are comparable in area to the major voids.
    
To conduct this test, similarly to the procedure described in Sect.~\ref{S:strong.B.in.voids}, we tiled the QS areas of the whole FOV into squares with an area similar to that of a coronal void. We then calculated the flux imbalance in each tile, and from all tiles we derived a histogram of (relative) flux imbalances for the given size of the tiles. As before, we allowed the tiles to overlap (in that each tile is shifted by 5\,Mm relative to the adjacent one). The procedure was carried out for tile sizes corresponding to both the smallest major void (diameter of 20\,Mm) and the largest one (77\,Mm).
    
The relative flux imbalance in the tiled QS region reaches up to 0.8 in the case of the small tiles (corresponding to a size of 20 Mm), but only up to 0.2 in the case of the larger tiles (77 Mm). This is illustrated by the histograms in Fig.\ \ref{distribution_imbalance_in_QS} and is consistent with the expectation that the larger the (QS) region under consideration the smaller the flux imbalance. The flux imbalance of the major voids lies mostly within the range covered by the histogram formed by the 77\,Mm tiles, except for the two smallest of the major voids, which, however, lie within the range covered by the histogram formed by the 20\,Mm tiles (Fig.\ \ref{distribution_imbalance_in_QS}).
    
This result leads us to conclude that the flux imbalance we find for the major voids is consistent with QS regions of comparable size. Hence, based on our data, we do not find a significant flux imbalance in the major voids, at least the flux imbalance in the major voids is fully consistent with the QS.

\subsection{Magnetic field extrapolation} \label{subsection_extrapolation}

\begin{figure}
\centering
    \resizebox{\hsize}{!}{\includegraphics{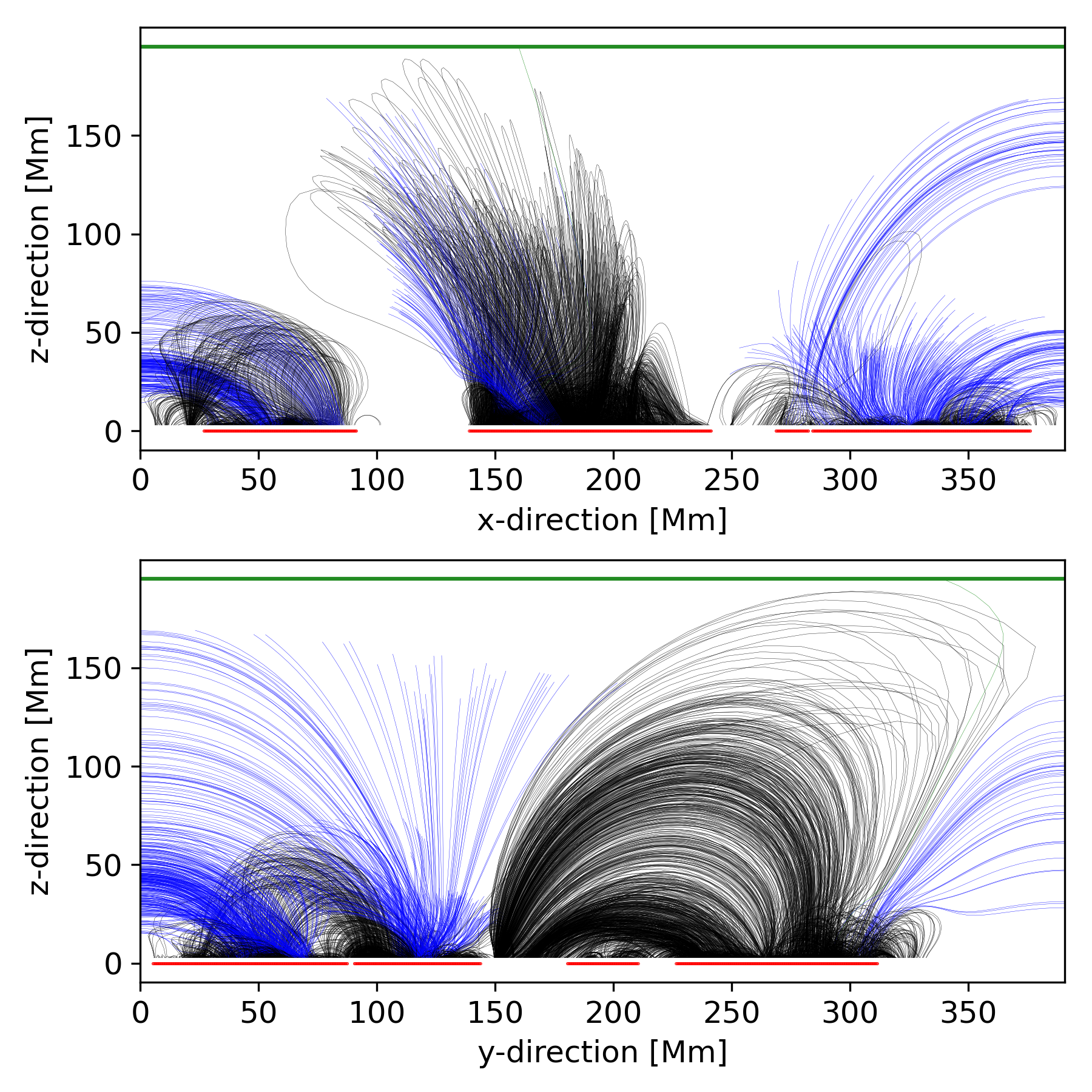}}
    \caption{Field lines traced within a potential field extrapolation. The field lines originate from a regular horizontal grid (with a 2-pixel grid scale) 2\,Mm above the photosphere, but only from grid cells located within coronal voids (although the other footpoint can end outside a void). Every tenth field line is displayed here and has been projected onto the $x-z$ and $y-z$ planes. Field lines that reach the upper boundary, leave the simulation box sideways, and close back to the photosphere are shown in green, blue, and black, respectively. The horizontal green line denotes the upper boundary of the simulation box. The red lines at the bottom of both panels indicate the positions of the coronal voids.}
     \label{field_lines_sideview}
\end{figure}

To better distinguish coronal voids from CHs, we computed a potential field using the original magnetogram as boundary conditions. The upper boundary of the extrapolations volume is placed at 195\,Mm, which corresponds to the length covered by 1023 horizontal pixels. Field lines reaching this upper boundary are considered to be open within the limitations of our extrapolation model. We solved the potential field equations with the help of a fast Fourier transform method as described in \cite{Alissandrakis1981_extrapolations}. Since the original HRT magnetogram is not globally flux-balanced, we applied a similar technique as in \cite{Seehafer1978boundarycond}, namely we extended the magnetogram by its three-point mirror images to ensure flux balance, and then applied the fast Fourier method.
   
We traced field lines starting from seed points at a height of 2\,Mm, but only above the voids (at every second horizontal grid point in both $x$ and $y$ directions) and then traced the field lines upwards. In total, nearly 75000 field lines were obtained this way.
    
Only 0.007\%\ of all traced field lines (i.e. five field lines in total) reach the upper simulation box boundary at 195\,Mm (see Fig.~\ref{field_lines_sideview}).
Most of the field lines are closed within one void or connect from the voids to the photosphere in QS areas outside the voids. There are no field lines connecting from one void to another void of opposite net magnetic polarity. Some field lines leave the simulation box sideways. However, because this behaviour may be influenced by the mirroring applied to the boundary conditions, we cannot draw conclusions about these field lines. This would require a magnetogram with a much larger FOV.

We took a closer look at the origin of field lines reaching the upper boundary at 195\,Mm (originating from the voids and the QS alike). Though field lines that reach this boundary are considered open in our simulation, they do not necessarily indicate real open fields because they could close outside the simulation box.
We placed seeds (at every fifth horizontal grid point covering the whole upper boundary) and traced field lines from this upper boundary downwards to the photosphere to investigate the connectivity of the `open' field lines.
More than 83\% of these open field lines originate from a strong flux concentration at $x{=}200$, $y{=}140$ in the middle of the FOV (see green contours in Fig.~\ref{EUI+PHI_overview}). 15\% are from QS fields in the upper right area on the magnetogram. Less than 2\% can be traced back to an area in the largest coronal void (at $x{=}160$, $y{=}280$ in Fig.~\ref{EUI+PHI_overview}). However, these field lines have their foot points in a stronger flux concentration that is next to an EUV-brightness surrounded by the void. Therefore, we conclude that the coronal voids are magnetically closed and hence distinct from CHs, which are typically dominated by open fields at coronal heights.

\section{Discussion and conclusions}
We report EUV-dark areas present in \euv\ images (sampling the wavelengths around 174\,\AA) of the QS. We called these dark areas `coronal voids'. They are outlined very well by an iso-contour of 75\% of the average QS intensity in the same 174\,\AA\ channel (Sect. \ref{S:define.voids}). The chromospheric emission originating in the Ly-$\alpha$ line in the voids is also reduced (Sect. \ref{section_on_lyman-alpha}). In our dataset we find a range of such voids with sizes comparable to and larger than a supergranular cell. 
An earlier study found dark threads projected against coronal loops at the limb \citep{DarkThreads_November+Koutchmy_1996}. It is, however, unclear whether and how those dark thread features might be related to the coronal voids that we study here, in particular because the dark threads appear to be significantly smaller than the coronal voids.
    
We started out with two hypotheses as to the origin of the reduced EUV emission in the coronal voids: (1) The photospheric magnetic field underlying the coronal voids is less strong and hence the magnetic heating there is reduced, leading to lower temperatures and densities and thus lower coronal emission as compared to the QS. (2) The voids are miniature versions of CHs with open field lines due to a significant flux imbalance in the photosphere.
    
We find that the mean absolute LOS magnetic field, |\blos|, inside the coronal voids is a factor of 1.3--1.6 lower than in the surrounding QS (Sect.\ \ref{S:B.in.voids}). Equally important, the major voids are almost free of the patches of stronger magnetic fields that are abundantly found in the QS network (Sect.\ \ref{S:strong.B.in.voids}). Furthermore, many regions of weak averaged magnetic field coincide with (small or large) coronal voids (Sect.\ \ref{S:weak.B.def.voids}). Taken together, this shows that there is less magnetic activity below and in the coronal voids.
This implies that there is a smaller flux of magnetic energy into the upper atmosphere of the voids, and hence the chromosphere and corona in such regions are heated significantly less than in the ambient QS. According to general scaling relations \cite[][]{1978ApJ...220..643R}, this then implies that both the temperature and the density, and consequently also the coronal emission, should be reduced \cite[][]{2020A&A...640A.119Z}.
    
We find some imbalance of the magnetic flux in the coronal voids. It is considerably lower than the imbalance found in CHs, but is still present (Sect.\ \ref{subsection_flux_imbalance}). Thus, one could argue that the reason for the reduced coronal emission in the voids is similar to the case for CHs: the flux imbalance could signal that some of the field is open such that energy is lost to the acceleration of gas away from the solar surface. Consequently, the energy available to heat the gas is reduced, and hence the coronal temperature, density, and emission are lower than in the QS. However, further tests showed that the flux imbalance in the voids is of the same order as comparably sized patches of QS. Hence, we conclude that the flux imbalance we find in the voids is not significant. Furthermore, a potential magnetic field extrapolation indicates that there are no open fields originating from within the void. As a consequence of both findings, we consider it unlikely that coronal voids are essentially small CHs.
    
To further confirm our conclusions, it would be helpful to add coronal spectroscopic measurements to new observations of more voids, that is to say, to check the Doppler shifts in the voids and compare them to the ambient QS. This will show if an outflow as found in CHs is present in the voids or not.  
Also, it would be highly profitable to perform more QS observations to get better statistics of the properties of coronal voids, the reduction in magnetic flux density associated with them, and the flux imbalance distribution. 
A study following the evolution of coronal voids on timescales from hours to days could provide insight into their stability and possibly their lifetimes. In addition, further datasets that have co-observations with other instruments, for example \aia,\ would allow for a study of the coronal voids at other coronal temperatures. 

In summary, via a comparison of EUI and SO/PHI data, we have uncovered the presence of coronal voids, particularly dark parts in the QS corona that are associated with weaker than normal magnetic flux densities. Consequently, a reduction in magnetic energy input into the coronal gas appears to be the major cause of these darker coronal regions. All in all, our results suggest that coronal voids are not just small CHs, but rather have a distinct physical origin and hence belong to a different class of coronal features.
    
\begin{acknowledgements}
Solar Orbiter is a space mission of international collaboration between ESA and NASA, operated by ESA. We are grateful to the ESA SOC and MOC teams for their support. The German contribution to SO/PHI is funded by the BMWi through DLR and by MPG central funds. The Spanish contribution is funded by AEI/MCIN/10.13039/501100011033/ and European Union “NextGenerationEU”/PRTR” (RTI2018-096886-C5, PID2021-125325OB-C5, PCI2022-135009-2, PCI2022-135029-2) and ERDF “A way of making Europe”; “Center of Excellence Severo Ochoa” awards to IAA-CSIC (SEV-2017-0709, CEX2021-001131-S); and a Ramón y Cajal fellowship awarded to DOS. The French contribution is funded by CNES.
The EUI instrument was built by CSL, IAS, MPS, MSSL/UCL, PMOD/WRC, ROB, LCF/IO with funding from the Belgian Federal Science Policy Office (BELSPO/PRODEX PEA 4000134088, 4000112292, 4000136424, and 4000134474); the Centre National d’Etudes Spatiales (CNES); the UK Space Agency (UKSA); the Bundesministerium für Wirtschaft und Energie (BMWi) through the Deutsches Zentrum für Luft- und Raumfahrt (DLR); and the Swiss Space Office (SSO). 
This work was supported by the International Max-Planck Research School (IMPRS) for Solar System Science at the TU Braunschweig. 
L.P.C. gratefully acknowledges funding by the European Union (ERC, ORIGIN, 101039844). Views and opinions expressed are however those of the author(s) only and do not necessarily reflect those of the European Union or the European Research Council. Neither the European Union nor the granting authority can be held responsible for them. 
The ROB team thanks the Belgian Federal Science Policy Office (BELSPO) for the provision of financial support in the framework of the PRODEX Programme of the European Space Agency (ESA) under contract numbers 4000112292, 4000134088, 4000136424, and 4000134474. 
D.M.L. is grateful to the Science Technology and Facilities Council for the award of an Ernest Rutherford Fellowship (ST/R003246/1). 
We thank Yajie Chen for his help with the field line tracing.
\end{acknowledgements}


\end{document}